\documentclass[floatfix,nofootinbib,showpacs,
showkeys,aps,prd,reprint]
{revtex4-1}

\usepackage{bm}
\usepackage{amssymb}
\usepackage{slashed}
\usepackage{graphicx}

\begin{document}

\title{Thermal nonlocal Nambu--Jona-Lasinio model in the real time formalism}

\author{M. Loewe$^{1}$, P. Morales$^{1,2}$ and C.
Villavicencio$^{1,3}$}
\affiliation{$^1$Facultad de F\'isica, Pontificia Universidad
Cat\'olica
de Chile, Casilla 306, Santiago 22, Chile\\
$^2$Department of Physics, Graduate School of Science,
The University of
Tokyo 7-3-1 Hongo, Bunkyo-ku, Tokyo 113-0033, Japan\\
$^3$Universidad Diego Portales,  Casilla 298-V, Santiago,
Chile}

\begin{abstract}
The real time formalism at finite temperature and chemical potential
for the nonlocal Nambu--Jona-Lasinio model is developed in the
presence of a Gaussian covariant regulator. 
We construct the most general thermal propagator, by means of the
spectral function. 
As a result, the model involves the propagation of massive
quasiparticles.
The appearance of complex poles is interpreted as a confinement signal,
and, in this case, we have unstable quasiparticles with a finite decay
width. 
An expression for the propagator along the critical
line, where complex poles start to appear, is also obtained.
A generalization to other covariant regulators is proposed.
\end{abstract}


\maketitle

The Nambu--Jona-Lasinio model (NJL) has been vastly considered for studying
nonperturbative aspects of QCD. 
Nowadays, it is mainly used to explore
finite temperature and density effects in the frame of the mean-field
approximation \cite{Klevansky:1992qe, Buballa:2003qv, Loewe:2008kh}. 
One of the big challenges in QCD is to understand the confinement
mechanism and the dynamics behind confinement.
Perturbative QCD cannot describe confinement and, 
although lattice QCD is able to reproduce successfully hadron
properties, like masses and coupling constants \cite{Lattice2009}, it
has problems when dealing with finite baryon chemical potential (the sign
problem).
However, there are effective models which include explicitly
confinement, as, for example, different versions of the bag model
\cite{DeFrancia:1996eh, DeFrancia:1998if, Fraga:2008be,
Hansen:2006ee},
Dyson-Schwinger models \cite{Roberts:1994dr, Blaschke:1998gk,
Blaschke:1999ab, Roberts:2000aa, Blaschke:2000gd}, 
or the Polyakov loop effective action coupled to Dyson-Schwinger or NJL
models
\cite{Meisinger:1995ih, Fukushima:2003fw, Mukherjee:2006hq,
Fukushima:2008wg}.

The nonlocal  NJL model (nNJL) is another attempt in this
direction \cite{Bowler:1994ir, Plant:1997jr,Radzhabov:2003hy,
GomezDumm:2006vz}. 
When the gluon degrees of freedom are integrated out in the QCD
action, a nonlocal quark action emerges and confinement should be
hidden there. 
The idea of the nNJL approach is to incorporate nonlocal
vertices through the presence of appropriate regulators.

Since the NJL model is nonrenormalizable, a momentum cutoff is needed
in order to handle the UV divergences. 
The applicability of the model is, therefore, restricted to energy
scales below the cutoff. 
Nonlocal extensions of the NJL model are designed to regularize the
model in such a way that UV
divergences are controlled, internal symmetries are
preserved, and quark confinement is incorporated.
The nNJL model has been extended to a finite temperature and density
scenario \cite{General:2000zx, GomezDumm:2001fz,
GomezDumm:2005hy, Blaschke:2007np, Contrera:2007wu, Hell:2008cc, Hell:2009by, Contrera:2009hk,
Contrera:2010kz,Radzhabov:2010dd, Horvatic:2010md} through the usual Matsubara
formalism \cite{
LeBellac, Das:1997gg}.
Here, nevertheless, the exact summation of Matsubara frequencies turns out to be
cumbersome, due to the complicated shape of the regulators. 
In most cases, it is necessary to cut the series at some order.

The idea of this work is to develop the finite temperature real time
formalism for the nNJL model.
In this way, we are able to calculate temperature corrections,
providing a physical picture in terms of quasiparticles.
On the one hand, those states with real masses can propagate freely in
the deconfined phase.
On the other hand, the existence of complex poles of the propagator
in the confined phase produces a strong damping avoiding the
propagation of such states.

Real time formalisms, as thermo-field dynamics or the
Schwinger-Keldysh formalism, can be constructed through the analytic
continuation of the Euclidean action \cite{Ojima:1981ma,
Matsumoto:1983by, Kobes:1984vb, Landsman:1986uw, LeBellac,
Das:1997gg, Khanna:2009zz}.
Those formalisms double the number of degrees of freedom, providing
the appearance of a $2\times 2$ matrix propagator $S_{ij}$. 
In any version of the NJL model in the mean-field approximation, the
gap equation corresponds to a one-loop self-consistent relation,
so we only need to find the $S_{11}$ component
\cite{Dolan:1973qd,Ebert:1992ag}. 
The other matrix propagator components start to appear at the two-loop
level.
In the construction of the $S_{11}$ propagator, the main ingredient
is given
by the spectral density function.
However, the construction of the spectral density function is not a
simple task in a nonlocal frame, especially when dealing with a
nontrivial analytic
structure of the propagator.

In this article, we will develop the real time formalism for the
Gaussian regulator in the nNJL model, which  can easily be extended to
 other kind of regulators.
As a result, we will get thermal propagators describing quasiparticles
with temperature- and chemical-potential--dependent masses.

\bigskip
This paper is organized as follows:
In Sec. \ref{nNJL}, we introduce the nNJL model in the mean-field approximation 
at finite temperature and chemical potential.
Section \ref{sect.poles} is devoted to the discussion of the analytical structure of 
the nNJL model in the mean-field approximation with a Gaussian regulator.
In Sec.  \ref{sect.real-time}, the real time formalism is presented for the general case, 
constructing, then, in Sec.
\ref{sect.propagators}, the corresponding real time propagator $S_{11}$ for the
case of a Gaussian regulator. 
We explain how to obtain the gap equation and the thermodynamical potential 
in Sec. \ref{sect.gap} and Sec. \ref{sect.Omega}, respectively.
Finally, in Sec. \ref{sect.conclusions}, we summarize our conclusions.

\section{$\mathrm{n}$NJL model in the mean-field approximation at finite
temperature and chemical potential}
\label{nNJL}

Nonlocal models with separable interactions have been considered
several times. 
In the context of NJL, confinement is introduced through the
inclusion of interacting nonlocal currents, which are extended
regularized versions of the usual local currents.

The $SU(2)_f\otimes SU(3)_c$ nNJL Euclidean Lagrangian is given by
\begin{equation}
 {\cal L}_E = \bar \psi(-i
\slashed{\partial}+m)\psi-\frac{G}{2}j_a(x)j_a(x),
\end{equation}
$m$ being the current quark  mass and $j_a(x)$ the nonlocal quark
currents. 
Here, we are using the metric $g_{\mu\nu}=\mathrm{diag}(1,1,1,1)$.

There are two schemes on the market to introduce nonlocal effects 
in terms of extended currents \cite{GomezDumm:2006vz} .
We will use the one based on instanton liquid models:
\begin{equation}
j_{a}(x) = \int  d^{4}y\; d^{4}z ~r(y-x)r(z-x)\bar{\psi
}(y)\Gamma
_{a}\psi(z),
\label{jnolocal}
\end{equation}
where $r(x-y)$ is the nonlocal regulator and $\Gamma_{a} =
(1,i\gamma_{5}\bm{\tau})$, $\bm{\tau}$ being the Pauli matrices.

Since we want to deal with mesonic degrees of freedom, we will follow
the standard bosonization  procedure.
This is realized  through the introduction of auxiliary scalar and
pseudoscalar fields.
By integrating out the quark fields, an equivalent partition
function, in terms of only bosonic degrees of freedom, is obtained:
\begin{equation}
 {\cal Z} = \int D\sigma D^3\pi  ~e^{-\Gamma[\sigma,\pi^a]}.
\label{Zboson}
\end{equation}
We proceed within the mean-field approximation, keeping the mean
values of the boson fields and neglecting their fluctuations.
In this way, the partition function in Eq. (\ref{Zboson}) turns out to
be ${\cal Z}\approx {\cal Z}_{\mathrm{MF}} =
e^{-\Gamma_{\mathrm{MF}}}$, where the
mean-field effective action in Euclidean momentum space is given by
\begin{equation}
\Gamma_{\mathrm{MF}}=V_{4}\left[\frac{\bar\sigma^2}{2G}
-2N_c \int 
\frac{d^4q_E}{(2\pi)^4}\mathrm{tr} \ln S_E^{-1}(q_E)
\right],
\label{SMF}
\end{equation}
$q_E=(\bm{q},q_4)$ being the four-momentum in Euclidean space and
  $V_4$
the four-dimensional volume.
The trace acts on the Euclidean Dirac
matrices $\gamma_E=(\bm{\gamma},i\gamma_0)$   of the effective quark
propagator
\begin{equation}
 S_E(q_E) =
\frac{-\slashed{q}_E+\Sigma({q_E}^2)}{{q_E}^2+\Sigma^2({q_E}^2)},
\label{prop-euclideo}
\end{equation}
where the running mass includes the Lorentz invariant regulator 
contribution in Euclidean momentum-space
\begin{equation}
 \Sigma({q_E}^2) = m+\bar\sigma r^2({q_E}^2),
\end{equation}
$\bar \sigma$ being the mean-field value of the scalar field.
The pseudoscalar mean-field value is zero due to isospin
symmetry. 
Chiral symmetry, on the other hand, is explicitly broken
through the
current quark masses and spontaneously broken by a nonvanishing
chiral condensate value.

Finite temperature ($T$) and chemical potential ($\mu$) effects are
introduced in the standard way through the Matsubara formalism
\cite{LeBellac}. 
As a result, in Eq. (\ref{SMF}), the four-dimensional volume, the
momentum, and the integral in momentum turn out to be
\begin{eqnarray}
 V_4 &\to& V/T, \\
q_4 &\to&  -q_{n},\\
\int \frac{dq_4}{2\pi}
&\to& T\sum_n,
\end{eqnarray}
where $q_n$ includes the Matsubara frequencies and the chemical
potential
\begin{equation}
 q_n\equiv (2n+1)\pi T+i\mu.
\end{equation}

The grand canonical thermodynamical potential is given by
$\Omega_{\mathrm{MF}}(\bar\sigma,T,\mu)=
(T/V)  \Gamma_{\mathrm{MF}}(\bar\sigma,T,\mu)$ \cite{Kapusta:1989tk}.
The value of $\bar\sigma$ is obtained then through the solutions
of $\partial
\Omega_{\mathrm{MF}}/\partial\bar\sigma=0$, giving rise to the
gap equation
\begin{equation}
\frac{\bar\sigma}{G}=
2 N_c T \sum_n\int\frac{d^3q}{(2\pi)^3}
r^2(q_n^2+\bm{q}^2)\mathrm{tr}S_E(\bm{q},-q_n).
\label{gap_Matsubara}
\end{equation}
This equation, which involves directly the propagator, will be our
starting point in order to apply the real time formalism.

\section{Poles of the Effective Quark Propagator with a Gaussian
regulator in Minkowski space}
\label{sect.poles}

Among the different kind of existent covariant regulators, we will
use here the Gaussian one due to its simplicity
\cite{Bowler:1994ir}. 
In Euclidean momentum space, the regulator  is given by
\begin{equation}
r({q_E}^{2}) = \exp\left( -\frac{{q_E}^{2}}{2\Lambda ^{2}}\right),
\end{equation}
where $\Lambda $ is a free parameter that has to be chosen from
phenomenological considerations.
Thus, the effective model depends on three parameters: the current
quark mass $m$, the effective coupling $G$, and the scale parameter
$\Lambda$, the last one being associated to the cutoff in the usual
NJL model by setting the regulator as $r=\theta(\Lambda^2-\bm{q}^2)$.
These parameters are fixed in order to get the physical values
of the pion mass $m_\pi=139$ MeV, the pion decay constant
$f_\pi=92.4$ MeV, and the chiral condensate
$-\langle\bar qq\rangle^{1/3} \simeq 200$--$260$ MeV
\cite{GomezDumm:2006vz}. 
Following \cite{GomezDumm:2001fz}, we will use two sets of parameters.
Set I is given by $\langle \bar qq\rangle =-(200$ MeV$)^3$, 
$m=10.5$ MeV, $G=50$
GeV$^{-1}$ and $\Lambda= 627$ MeV. 
Set II is given by $\langle \bar qq\rangle =-(220$ MeV$)^3$,
$m=7.7$ MeV, $G=30$ and $\Lambda= 760$ MeV.

\begin{figure}
 \includegraphics[scale=.6]{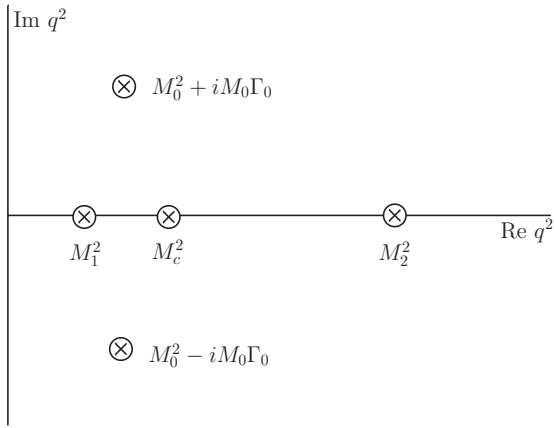}
\caption{Schematic description of the poles of the propagator in
Minkowski space. $M_1$ and $M_2$ correspond to the deconfined masses,
$M_c$ is the critical mass and $M_0$ and $\Gamma_0$ are the mass and
the decay width in the confined phase, respectively.}
\label{poles}
\end{figure}

To start our analysis in the real time formalism, let us explore more in
detail the analytic structure of the propagator by setting $q_4=iq_0$.
The Euclidean propagator in Eq. (\ref{prop-euclideo}) turns out to be
$S_E\to iS_0$, where the propagator in Minkowski space is defined as
\begin{equation}
S_0(q)=i\frac{\slashed{q}+\Sigma(-q^2)}
{q^2-\Sigma^2(-q^2)},
\label{S0}
\end{equation}
and where now $q=(q_0,\bm{q})$ and
$\gamma=(\gamma_0,\bm{\gamma})$, with the metric
$g_{\mu\nu}=\mathrm{diag}(1,-1,-1,-1)$.
This propagator, as shown in Fig. \ref{poles}, presents three
different kind of poles which depend on  $\bar\sigma$. 
For Low values of $\bar\sigma$, there are two real poles associated
to real masses that we denote as $M_1$ and $M_2$. 
In this case we will say that the system is in the deconfined phase
since we have freely propagating quasiparticle states.
As $\bar\sigma$ grows up to a certain critical value,
the two masses join into a single real mass $M_c$. 
For higher values of $\bar\sigma$, this critical mass splits into
complex poles.
There exist an infinite number of those poles for the Gaussian regulator.
Nevertheless, it can be shown
that the only relevant poles correspond to the first pair
\cite{GomezDumm:2001fz}, where  the real part is associated  with a mass 
$M_0$ and the imaginary part is related to a decay width factor
$\Gamma_0$. 
The other complex poles involved have considerable higher values for
the masses and decay widths, and these masses are not continuously connected
with the critical mass $M_c$.
The appearance of complex poles is interpreted as a signal of
confinement \cite{Bowler:1994ir, Plant:1997jr}, since the corresponding 
quasiparticles do not propagate freely.
As soon as the poles becomes complex, we will be in the {\it confined phase}.

\begin{figure}
\includegraphics[scale=0.66]{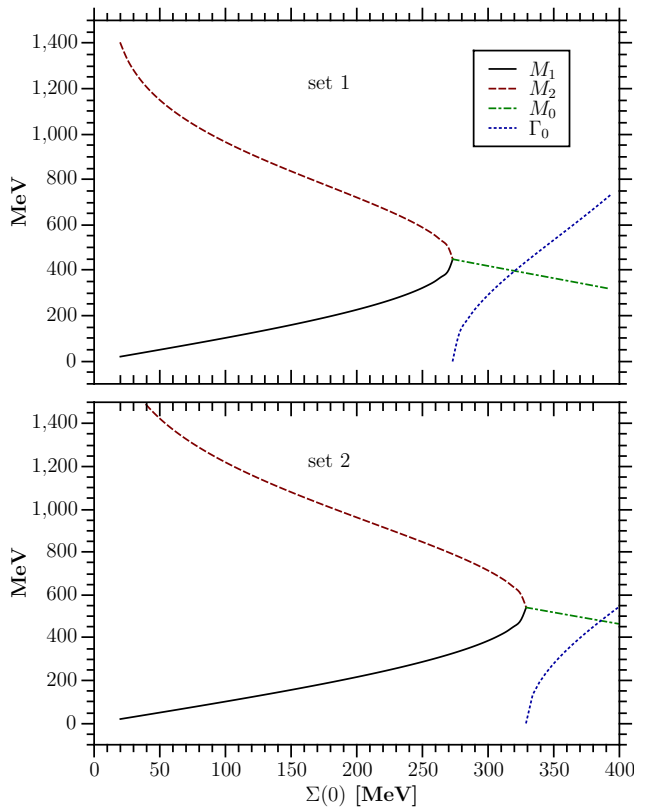}
\caption{Poles of the propagator as a function of the self-energy at
zero 4th-momentum for set I (upper graph) and set II (lower
graph). Defining the complex poles as $q^2=M_0^2\pm iM_0\Gamma_0$, the
solid, dashed, and dot-dashed lines correspond to the masses, and
the dotted line corresponds to the decay width.}
\label{polesvsS0}
\end{figure}

The poles of the propagator in Eq. (\ref{S0}) are plotted in Fig.
\ref{polesvsS0} for both sets of parameters as a
function of the self-energy at zero 4th-momentum
$\Sigma(0)=m+\bar\sigma$.
When $\Sigma(0)=\Sigma_c$, there is only one positive real
pole $q^2=M_c^2$, which is determined from the
condition $\partial_{q^2} [q^2-\Sigma^2(-q^2)]=0$, giving
\begin{eqnarray}
 \Sigma_c &=& m+(M_c-m)e^{-M_c^2/\Lambda^2},\\
M_c &=& \frac{1}{2}\left(m+\sqrt{m^2+2\Lambda^2}\right).
\label{Mc}
\end{eqnarray}
 The corresponding values are $\Sigma_c = 273$ MeV for set I and
$\Sigma_c = 329$ MeV for set II.
The previous analysis can be easily verified by plotting
$q^2-\Sigma^2(-q^2)$ as a function of $q^2$.

The values of $\Sigma(0)$ at zero temperature and chemical
potential are $\bar\Sigma = 350$ MeV for set I and $\bar\Sigma
= 300$ MeV
for set II, respectively. 
The tendency for the $\bar\sigma$ parameter is to decrease as temperature 
and chemical potential increase. 
Then, if at zero temperature and chemical potential the corresponding poles are real, 
they  will still be real at finite $T$ and $\mu$.
On the other hand, if the poles are complex at zero temperature and chemical potential, 
they will become real at some finite values of $T$ and $\mu$.  
As $\bar\Sigma>\Sigma_c$ in set I and $\bar\Sigma<\Sigma_c$ in
set II,
they are called {\it confining} and {\it nonconfining} sets,
respectively.

Now we proceed to extend this formalism to a finite temperature and density
scenario in the frame of real time formalism.

\section{Real time formalism}
\label{sect.real-time}

It is well-known that when formulating quantum field theory at finite
temperature in the real time formalism, we have to double the number
of degrees
of freedom \cite{Ojima:1981ma,
Matsumoto:1983by, Kobes:1984vb, Landsman:1986uw, LeBellac,
Das:1997gg, Khanna:2009zz}. 
The new fields that appear are called thermal ghosts. 
As a consequence, the new thermal propagators are given by $2\times
2$ matrices $S_{ij}$.
However, for one-loop calculations, we only need the term $S_{11}$.
The other components of the thermal matrix propagator participate
only in higher loop calculations, since the thermal ghosts do not
couple directly to external physical lines.
The treatment of the gap equation, in a self-consistent way, is
equivalent to a one-loop calculation.
The same happens with the calculation of the chiral condensate and the
number density.
For our purpose, we only need to obtain the $S_{11}$
component of the propagator matrix.

\begin{figure}
\includegraphics[scale=0.46]{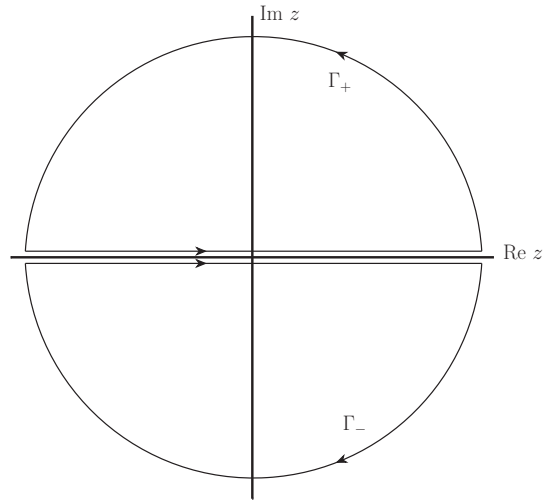}
\caption{Integration path in the definition of $S_\pm$.
In both paths the integral is taken along the real axis, but the
curve is closed through the upper half plane in $\Gamma_+$ and
through the lower half plane in $\Gamma_-$. }
\label{Gama_pm}
\end{figure}

Following the construction of \cite{Dolan:1973qd},
the general $S_{11}$  propagator is obtained in terms of the spectral
density function (SDF):
\begin{equation}
S_{11}(q) = \int \frac{dk_0}{2\pi i}
\frac{\rho(k_0,\bm{q})}{k_0-q_0-i\epsilon}-
n_F(q_0-\mu)\rho(q).
\label{S11}
\end{equation}
The SDF is related to the real time propagator in Eq. (\ref{S0}) through
\begin{equation}
 S_0(iq_n,\bm{q})=\int \frac{dk_0}{2\pi
i}\frac{\rho(k_0,\bm{q})}{k_0-iq_n},
\label{SDF}
\end{equation}
where the connection between both formalisms,
real time and imaginary time, 
is realized through the analytic
extension of $iq_n\to z$. 
 What we need now is to obtain the SDF.
In the case of free particles, the SDF can be gotten from
the relation $\rho = S_0(q_0+i\epsilon,\bm{q})-
S_0(q_0-i\epsilon,\bm{q})$.
However, in this case where, we have a nontrivial propagator, we
need a more general prescription to extract the SDF.
This can be achieved by defining
 \begin{equation}
S_{\pm}(q)\equiv\pm \oint_{\Gamma^{\pm}} \frac{dz}{2\pi i}
\frac{S_0(z\mp
i\epsilon,\bm{q})}{z-q_0 +i\epsilon},
\label{Spm}
\end{equation}
where the integration path is shown in Fig. \ref{Gama_pm}.
%%%%%%%%%%%%%%%%%%%%%
%%%%%%%%%%%%%%%%%%%%%%%
With this definition, the SDF can be written as
\begin{equation}
 \rho(q)=S_+(q)-S_-(q).
\end{equation}

In the special case of free fermions with mass $M$, the corresponding
spectral density function will be 
\begin{equation}
 \rho_{\mathrm{free}}(q) =  2\pi ~\mathrm{sign}(q_0)
 (\slashed{q}+M)\delta(q^2-M^2).
\end{equation}
By replacing $ \rho_{\mathrm{free}} $ in Eq. (\ref{S11}), we find the
 Dolan-Jackiw  (DJ) propagator
\cite{Dolan:1973qd}
 \begin{eqnarray}
 S_{\mathrm{DJ}}(q;M) &=&
(\slashed{q}+M)\bigg[\frac{i}{q^2-M^2+i\epsilon}
\nonumber\\&& 
-2\pi N(q_0)(\slashed{q}+M)\delta(q^2-M^2)
\bigg],
\label{SDJ}
\end{eqnarray}
 where the function $N(q_0)$ is defined in terms of the Fermi-Dirac
distribution $n_F(q_0)=(e^{q_0/T}+1)^{-1}$ and reads
\begin{equation}
 N(q_0)=\theta(q_0)n_F(q_0-\mu)+\theta(-q_0)n_F(\mu-q_0).
\end{equation}
Note that we make a distinction here between the usual case where a
particle is propagating freely and a more general situation where we
have dressed propagators.

Through this procedure, the SDF can be obtained even if the regulator
has a cut in the real axis. 
Consequently, the Euclidean gap equation, written  in Eq. (\ref{gap_Matsubara}), 
will take another form in the real time formalism 
by introducing the following replacements:
\begin{eqnarray}
 q_n &\to& -iq_0 \nonumber,\\
T\sum_n &\to& -i\int\frac{dq_0}{2\pi}\nonumber,\\
S_E(\bm{q},-q_n)&\to& iS_{11}(q).
\label{MatMink}
\end{eqnarray}
Now, we need to construct the dressed $S_{11}$ propagators.
Notice that, in the real time formalism, we are able to separate immediately 
the finite-temperature- 
and chemical-potential--dependent terms from the $T,\mu=0$
contributions.

\section{Dressed propagators for the Gaussian regulator}
\label{sect.propagators}

Following the procedure described above, we will start from
the deconfined region, where there are two simple real poles.
The corresponding $S_{11}$ propagator is
\begin{equation}
 S_{11}^{\mathrm{dec}}(q) =
Z(M_1) S_{\mathrm{DJ}}(q;M_1)
+Z(M_2)S_{\mathrm{DJ}}(q;M_2),
\label{Sdec}
\end{equation}
$S_{\mathrm{DJ}}(M)$  being the DJ propagator with mass $M$ 
 and where $Z(M)$ is the field-strength renormalization
constant, defined as 
\begin{eqnarray}
 Z(M) &=&
\left[\partial_{M^2}\left\{M^2-\Sigma^2(-M^2)\right\}\right]^{-1}
\nonumber\\
&=& \left[1-2M(m-M)\right]^{-1}.
\label{Z}
\end{eqnarray}
In the last step, we used the pole relation $M=\Sigma(-M^2)$. 
Now, this propagator describes the quasiparticles involved, and the nonlocal
interaction is enclosed in the effective masses and the
field-strength renormalization constants.

The critical case that separates the real  from the  complex
pole regions is a little  bit different from the case we already discussed.
As we can see from Fig. \ref{polesvsS0}, the two real poles converge into a 
single real pole at $\Sigma(0)=\Sigma_c$. 
However, this particular case corresponds to a second-order
pole.
 In addition, the new relation 				
$\partial_{q^2}[q^2-\Sigma^2(-q^2)]=0$ must also be satisfied. 
Following the Cauchy theorem, by expanding in a 
Laurent series and using the general expression that provides SDF, we find 
the critical propagator
\begin{equation}
S_{11}^{\mathrm{crit}}(q)=
\left(Z_c+Z'_c\frac{\partial}{\partial M_c^2}\right)
S_{\mathrm{DJ}}(q;M_c) ,
\end{equation}
 with the constants 
\begin{eqnarray}
 Z_c &=&\frac{2M_c(4 M_{c}-3m)}{3(2 M_{c}-m)^2}, \\
Z'_c &=&-\frac{4M_c^2(M_c-m)}{2M_{c} -m},
\label{Sc}
\end{eqnarray}
and where  $M_c$ was already defined in Eq. (\ref{Mc}).

In  both cases described above (deconfined and critical), 
the finite temperature and
chemical potential effects appear separated from the $T,\mu=0$ terms.
Notice that the finite temperature and chemical potential terms 
are on mass shell due to the delta function in the propagators. 
This fact does not imply any difficulty when integrating in $q_0$. 
The case of complex poles needs more attention.
The spectral density function in this case is not proportional to a delta function, but
it is given by a Breit-Wigner distribution:
\begin{equation}
\rho_{\mathrm{conf}}(q) = 
\frac{M_0\Gamma_0 (A_++A_-)-i(q^2-M_0^2)(A_+-A_-)}
{(q^2-M_0^2)^2+M_0^2\Gamma_0^2},
\end{equation}
with  
\begin{equation}
A_\pm(q) \equiv \frac{2Z(M_\pm)}{\sqrt{\bm{q}^2+M_\pm^2}}
\left[q_0\left(\slashed{q}+M_\pm\right) 
-\gamma_0\left( q^2-M_\pm^2\right) \right],
\end{equation}
and where
\begin{equation}
M_\pm \equiv \sqrt{M_0^2\pm iM_0\Gamma_0}
\label{Mpm}
\end{equation}
are the first complex solutions of the equation
$M = \Sigma(-M^2)$.
$Z(M)$ was already defined in Eq. (\ref{Z}).
The $S_{11}^{\mathrm{conf}}$ propagator can be easily calculated from 
Eq. (\ref{S11}), 
but we will skip this in order to avoid long expressions.
In fact,  the full  $S_{11}^{\mathrm{conf}}$ will not be necessary 
for the rest of this article.

\bigskip

In what follows, we are interested to separate the thermal and density
effects.
 Then, we can write the propagator as
\begin{equation}
  S_{11}(q) = S_0(q) + \tilde S(q;T,\mu),
  \label{S11-2}
\end{equation}
where $\tilde S(q;0,0)=0$. In the case of the deconfined and critical
propagators, the $\tilde S$ term can be easily identified since it is
proportional to $N(q_0)$.

For the confined phase, 
we need to find  a procedure to calculate integrals   
in the complex energy plane that involve the confined propagator,
keeping in mind that when $T$ and $\mu$ vanish 
we have to recover the $S_0$ propagator inside the integral.
The finite temperature and chemical potential contribution
must be  an analytic function.
We can see from the general expression of the $S_{11}$ propagator that 
the last term in Eq. (\ref{S11}) contains all the thermal and 
density information.
Nevertheless, it does not vanish when $T,\mu \to 0$, but  $ n_F(q_0-\mu)\to\theta(-q_0)$, 
which is not an analytic function.
However, we can integrate the last term in Eq. (\ref{S11}), which is analytic, 
proceeding then to remove constants independent of $T$ and $\mu$.
Such constants then can be joined into the rest of the integral.
This can be
summarized by writing 
\begin{equation}
\tilde
S_{\mathrm{conf}}(q)=-n_F(q_0-\mu)\rho_{\mathrm{conf}}
(q)+C\delta^ { 4 }(q) , 
\label{Sconf}
\end{equation}
where $C$ is a divergent factor that will be fixed after integration
in order to produce a vanishing function when $T,\mu \to 0$, being
 absorbed into the $T,\mu=0$ contribution. 
Now we proceed to calculate these integrals in order to obtain the gap
equation, the thermodynamic potential, and all the other relevant
quantities.

\section{Gap equation}
\label{sect.gap}

\begin{figure}
 \includegraphics[scale=.4]{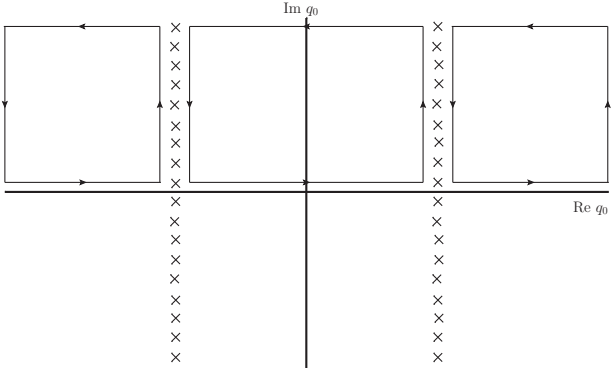}
\caption{Integration path for the thermal part of the gap equation in
the confined phase. 
The crosses represent the poles of the
Fermi-Dirac distribution $n_F(q_0\pm\mu)$.}
\label{path_gap_conf}
\end{figure}

The interesting quantities we want to calculate, such as the 
 thermodynamical potential, chiral condensate, number density,
susceptibilities, etc., require  momentum integration. 
In particular,
the thermodynamical potential at zero temperature and chemical
potential has to be calculated in the Euclidean formulation. 
However, the
thermal and density corrections can be handled in the frame of the
real time formalism.

The Gaussian regulator is constructed with the aim of regularizing UV
divergences in the Euclidean formulation.
Nevertheless, this regulator induces a divergence  in Minkowski space
when integrating in $q_0$.
The finite temperature and chemical potential part of the dressed 
propagators $\tilde S$, for the deconfined and critical case, involves
a delta function which has the effect of evaluating the regulator at
the effective masses $r(-q^2)\to r(-M^2)$.
The zero temperature and chemical potential term in the dressed 
propagator $S_0$ can be Wick rotated, turning back to Euclidean space,
where the regulator produces finite  integrals. 

\bigskip

We start first with the gap equation, which gives us the value of
$\bar\sigma$.
Following the previous section, we apply the replacements described in
Eq.
(\ref{MatMink}) to the gap equation in (\ref{gap_Matsubara}), obtaining
\begin{equation}
 \bar\sigma= 2N_c\int\frac{d^4 q}{(2\pi)^4}
r^2(-q^2)\mathrm{tr}S_{11}(q).
\end{equation}
Now, expressing the dressed propagator in the form of Eq.
(\ref{S11-2}), the gap equation turns out  to be 
\begin{equation}
 \frac{\partial\Omega_{\mathrm{MF}}}{\partial \bar\sigma} =
g_0(\bar\sigma)+\tilde g (\bar\sigma,T,\mu)=0,
\end{equation}
where turning back to
Euclidean space, the $T,\mu=0$ contribution gives
\begin{equation}
 g_0(\bar\sigma)=
\frac{\bar\sigma}{G}
-\frac{N_c}{\pi^2}\int_0^\infty 
dp p^3
\frac{r^2(p^2)\Sigma(p^2)}{p^2+\Sigma^2(p^2)},
\label{g_0}
\end{equation}
where $p=\sqrt{q_E^2}$ and the angular integral has already 
been done.
The finite $T,\mu$ contribution in Minkowski space  gives 
\begin{equation}
 \tilde g(\bar\sigma,T,\mu) =
-2N_c\int\frac{d^4q}{(2\pi)^4}r^2(-q^2)\mathrm{tr}\tilde S(q;T,\mu).
\label{tildeg}
\end{equation}
Notice that the $\bar\sigma$ variable and the different
mass terms are related through  $M=\Sigma(-M^2)$
or, more explicitly,
\begin{equation}
 \bar\sigma = (M-m)e^{-M^2/\Lambda^2},
\label{sigma-M}
\end{equation}
where $M$ stands for $M_{1,2}$ in the deconfined phase, and $M_c$ in
the critical phase. 
In the confined phase, $M$ stands for all
the complex poles in the confined phase,$M_\pm$ in Eq.
(\ref{Mpm})  being  the numerically relevant ones.

For the deconfined and critical phases, the integration can be done in a 
straightforward way.
However, 
as we mentioned in the last section,
the situation is not as simple when we try to integrate the
thermal part of the  propagator given in Eq. (\ref{Sconf}).
In Minkowski space,
the regulator in the gap equation diverges in some regions of the 
complex $q_0$ plane.
However, we will see that finally all the divergent terms do not depend
on temperature and chemical potential 
and, therefore,  can be removed, 
as will be explained in the last section.
Replacing $\tilde S$ from Eq. (\ref{Sconf})
into Eq. (\ref{tildeg}),
we get
\begin{eqnarray}
 \tilde g_\mathrm{conf} &=& 
N_c\int\frac{d^4q}{(2\pi)^4}r^2(-q^2)\left[
n_F(q_0-\mu)+n_F(q_0+\mu)\right]\nonumber\\
&&\times \mathrm{tr}\rho_\mathrm{conf}(q) 
- \frac{2N_c C }{(2\pi)^4},
\label{tildegconf}
\end{eqnarray}
where, in the last equation, we have used the fact that 
$\mathrm{tr}\rho_{\mathrm{conf}}$ is an odd function of $q_0$.
Now we integrate along the path shown in Fig.
\ref{path_gap_conf}.
If we consider only the two first poles of the propagator, $M_\pm$,
the poles inside the closed path will be $\sqrt{\bm{q}^2+M_+^2}$ 
and $-\sqrt{\bm{q}^2+M_-^2}$. 
The same results can be obtained if we integrate in the lower half
plane.
The integrand vanishes along the upper line ($\mathrm{Im}~q_0\to\infty$), 
and the contribution from the lines
surrounding the poles from the Fermi-Dirac factor (crosses) cancel
each other. 
The sum of the left and right straight lines ($\mathrm{Re}
~q_0\to\mp\infty$) leaves a divergent contribution to the integral 
which, however, is independent of temperature and chemical potential.
Therefore, this term, which is independent of $T$ and $\mu$, is 
canceled
with the constant factor $C$ in Eq.  (\ref{tildegconf}).

The deconfined and critical gap equations are obtained immediately, due to
the presence of the delta function in the propagators.
The resulting expressions for $\tilde g$ are
\begin{widetext}
\begin{eqnarray}
 \tilde g(\bar\sigma,T,\mu)
&=& \frac{4N_c}{\pi^2}\sum_{M}Z(M)
\int_0^\infty \frac{dk k^2}{2E}
r^2(-M^2)M\left[n_F(E-\mu)+n_F(E+\mu)\right],
\label{tildeg_conf_deconf}
\\
  \tilde g_{\mathrm{crit}}(\bar\sigma,T,\mu)
&=& \frac{4N_c}{\pi^2}\left(
Z_c+Z'_c \frac{\partial}{\partial M_c^2}\right)
\int_0^\infty \frac{dk k^2}{2E_c}
r^2(-M_c^2)M_c\left[n_F(E_c-\mu)+n_F(E_c+\mu)\right],
\\ &&\nonumber
\\ &&\nonumber
\\ &&\nonumber
\end{eqnarray} 
\end{widetext}
where $k=|\bm{q}|$, $E_c=\sqrt{k^2+M_c^2}$, $E=\sqrt{k^2+M^2}$, and
where the sum in the first equation is performed over all the mass
poles involved in the confined phase or the
deconfined phase.
For the case of a complex mass term,
the gap equation is real and can be easily written in terms of
real components.

\begin{figure}
 \includegraphics[scale=.6]{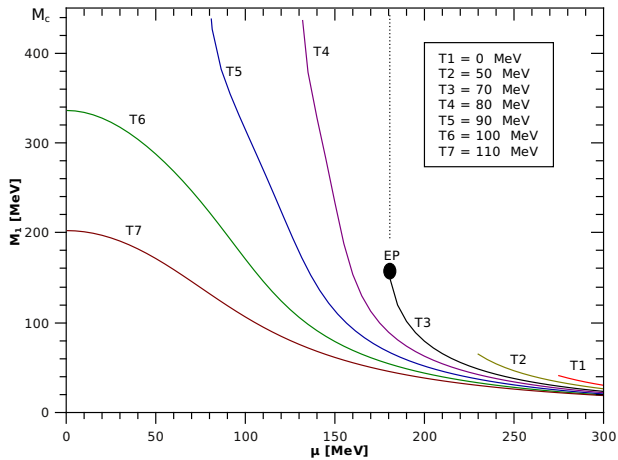}
\caption{$M_1$ as a function of the chemical potential for different
values of the temperature. The end point, where the first-order
phase transition turns to a crossover, occurs about $(T,\mu)\approx
(70, 180)$MeV.}
\label{plotM1}
\end{figure}

As an example, we calculate the behavior of $M_1$ in terms of the
temperature and chemical potential. 
This mass term will produce the main contribution to the dynamics 
in the deconfined phase because it is the lower one.
The other mass term $M_2$ turns out to be relevant only for values 
near the critical mass $M_c$.
Figure \ref{plotM1} shows the evolution of the mass $M_1$ as a function 
of the chemical
potential for different values of the temperature.
The transition observed at low temperature is obtained 
by analyzing the minimum of the thermodynamical potential as a function of 
$\bar\sigma$.
In the next section, we will show how to calculate the thermodynamical 
potential in the real time formalism through the gap equation.

\section{Thermodynamical potential}
\label{sect.Omega}

The inverse propagator in the real time
formalism does not carry information about temperature and density
\cite{Dolan:1973qd}. 
This can be seen in the case of a free fermion, whose
thermal propagator is the Dolan-Jackiw
one in Eq. (\ref{SDJ}), 
where the thermal information is enclosed in the on mass shell term.
However, the inverse of the full propagator, $-i(\slashed{q}-M)$,
 does not have any information on temperature.
Nevertheless, once we have obtained the expression for the gap equation, we can
reintegrate it, getting 
\begin{equation}
 \Omega_{\mathrm{MF}} =\int
\left(\frac{d\Omega_{\mathrm{MF}}}{d\bar\sigma}\right)d\bar\sigma
+\mathrm{const,}
\end{equation}
 where the constant is independent of $\bar\sigma$ and has to be
chosen  to regularize the thermodynamical potential.
In this way, as we did with the gap equation, we can separate the
zero and the nonzero contributions of the temperature and chemical
potential by defining
\begin{equation}
 \Omega_{\mathrm{MF}}(\bar\sigma;T,\mu)=
\Omega_0(\bar\sigma) +\tilde\Omega(\bar\sigma;T,\mu),
\end{equation}
 where the temperature and chemical potential contribution part
obeys  the relation $\tilde\Omega(\bar\sigma;0,0)=0$. 
 The term independent of the thermodynamical variables $\Omega_0$
can be obtained directly by integrating $g_0$ in Eq. (\ref{g_0}),
giving, as a result,
\begin{equation}
 \Omega_0(\bar\sigma)=
\frac{\bar\sigma^2}{2G}
-\frac{N_c}{2\pi^2}\int_0^\infty dp p^3 \ln [p^2+\Sigma^2(p^2)].
\end{equation}
The finite temperature and chemical potential expressions for the gap
equation, however, are not defined in terms of $\bar\sigma$, but in terms of
 the effective masses and the decay constant.
This relation is given in Eq. (\ref{sigma-M}), obtaining
\begin{equation}
 d\bar\sigma = Z^{-1}(M)e^{-M^2/\Lambda^2} dM,
\end{equation}
where we have used Eq. (\ref{Z}).
Proceeding with the integration, the temperature- and 
chemical-potential--dependent contribution to the thermodynamical potential 
will be
 \begin{eqnarray}
 && \tilde\Omega(\bar\sigma;T,\mu)=
-\frac{4N_c}{\pi^2}\sum_M\int_0^\infty dk k^2 ~T
\nonumber\\
&& \qquad\times\left[\ln\left(
1+e^{-(E-\mu)/T}\right)+ \mu\to-\mu
\right] ,
 \end{eqnarray}
where the sum, like in Eq. (\ref{tildeg_conf_deconf}), is performed
for all the poles involved.
Both phases will be continuously connected at $M=M_c$.
This is the familiar expression for the  thermodynamical potential.

The thermodynamical potential  with complex masses can be described in
terms of real components.
Considering only the numerically relevant poles $M_\pm$,
the energy term can be written as
\begin{equation}
 E_\pm = \omega \pm
i\frac{M_0\Gamma_0}{2\omega},
\end{equation}
with
\begin{equation}
 \omega =\sqrt{\frac{1}{2}\left[
k^2+M_0^2+\sqrt{(k^2+M_0^2)^2+M_0^2\Gamma_0^2}
\right]} .
\end{equation}
With these definitions, we can write the finite $T,\mu$ contribution
to the thermodynamical potential in the confined phase as
\begin{eqnarray}
 \tilde\Omega_{\mathrm{conf}}(\bar\sigma;T,\mu)&=&
-\frac{4N_c}{\pi^2}\int_0^\infty dk k^2 T
\nonumber\\&&
\times\left[\ln (1+B_-)+\ln(1+B_+)\right],
\end{eqnarray}
with the functions $B_\pm$ defined as
 \begin{equation}
B_\pm=
2\cos \left(\frac{M_0\Gamma_0}{2T\omega}\right)e^{-(\omega\pm \mu)/T}
+e^{-2(\omega\pm\mu)/T}.
\label{Bpm}
 \end{equation}
Our construction was compared with \cite{GomezDumm:2001fz}, 
obtaining the same results.
The cosine term in Eq. (\ref{Bpm}) can produce an unstable  
thermodynamical potential \cite{Blaschke:1999ab}. 
However, the inclusion of the Polyakov loop should fix this problem
\cite{Blaschke:2007np, Radzhabov:2010dd}.
Note that now, we can also obtain other relevant quantities
directly from the thermodynamical potential
already calculated in the real time formalism.
The quark number density function is obtained by taking the 
derivative with respect to the chemical potential: 
$n=-\partial\Omega/\partial\mu$.
The chiral condensate is calculated through the derivative of
the regulated thermodynamic potential with respect to the mass: 
$\langle \bar qq\rangle =\partial\Omega^r/\partial m$.
Here, the current mass is related to the effective mass terms through 
Eq. (\ref{sigma-M}), and the partial derivative is done by taking 
$\bar\sigma$ and $\Lambda$ as constants.

\section{Conclusions}
\label{sect.conclusions}

In the present work, we have constructed the real time formalism for the 
nonlocal Nambu--Jona-Lasinio model at finite temperature and chemical potential, 
in the particular case where the nonlocal term is given by a Gaussian regulator.
Following  the general construction of the $S_{11}$ propagator 
through the spectral density function \cite{LeBellac,Dolan:1973qd}, 
we generalize the procedure to get the spectral density function.
This generalization allows us to deal with dressed propagators 
whose analytical structure includes complex poles.
With this, we obtain different propagators if the quasiparticles are deconfined
(real poles), confined (complex poles), and also the critical case which separates both
regimes.
Once we have obtained the real time propagators, we find the gap equation which provides the 
value of $\bar\sigma$ and, consequently, the value of the quasiparticles effective mass terms.
The thermodynamical potential is obtained from the gap equation, and it is
given in terms of simple  expressions as a function of the effective mass terms. 
We verified that our results coincide with the ones reported in 
\cite{GomezDumm:2001fz}.

This procedure gives an intuitive phenomenological description
in terms of quasiparticles.
Its generalization to other regulators is, in principle, straightforward.
In fact, our prescription for obtaining the spectral density function is well-defined,
even if the propagator presents a cut along the real axis, 
which is the case for some type of Lorentzian regulator with fractional power.

\acknowledgments
The authors acknowledge support from FONDECYT under Grant No.
1095217. 
M.L. acknowledges also support from the Proyecto Anillos ACT119.
We thank N. N. Scoccola for suggesting this problem and for fruitful discussions.
We also thank D. Gomez-Dumm  and D. Blaschke for discussions
and criticisms.

\end{document}